\documentclass[aps,prd,twocolumn,letterpaper,superscriptaddress, amsmath, amssymb]{revtex4}

\usepackage{graphicx}
\usepackage{amsmath,amssymb}

\begin{document}


\title{New Sensitivity to Solar WIMP Annihilation using Low-Energy Neutrinos}

\author{Carsten Rott}
\affiliation{Center for Cosmology and AstroParticle Physics (CCAPP), Ohio State University, Columbus, OH 43210, USA}
\affiliation{Department of Physics, Ohio State University, Columbus, OH 43210, USA}
\affiliation{Department of Physics, Sungkyunkwan University, Suwon 440-746, Korea}

\author{Jennifer Siegal-Gaskins}
\affiliation{Einstein Fellow -- California Institute of Technology, Pasadena, CA 91125, USA}

\author{John F. Beacom}
\affiliation{Center for Cosmology and AstroParticle Physics (CCAPP), Ohio State University, Columbus, OH 43210, USA}
\affiliation{Department of Physics, Ohio State University, Columbus, OH 43210, USA}
\affiliation{Department of Astronomy, Ohio State University, Columbus, OH 43210, USA}

\date{5 September 2013}

\begin{abstract}
Dark matter particles captured by the Sun through scattering may annihilate and produce neutrinos, which escape.  Current searches are for the few high-energy neutrinos produced in the prompt decays of some final states.  We show that interactions in the solar medium lead to a large number of pions for nearly all final states.  Positive pions and muons decay at rest, producing low-energy neutrinos with known spectra, including $\bar{\nu}_e$ through neutrino mixing.  We demonstrate that Super-Kamiokande can thereby provide a new probe of the spin-dependent WIMP-proton cross section.  Compared to other methods, the sensitivity is competitive and the uncertainties are complementary.
\end{abstract}


\maketitle


\section{Introduction}
If dark matter is a thermal relic of the early universe, then its self-annihilation cross section is revealed by its present mass density.  To match observations, the required cross section, averaged over relative velocities, is $\langle \sigma_A v \rangle = (5.2 - 2.2) \times 10^{-26}$~cm$^{3}$/s, as a function of increasing mass, $m_\chi$~\cite{Steigman:2012nb}.  This indicates a weakly interacting massive particle (WIMP; denoted $\chi$)~\cite{Jungman:1995df, Bertone:2004pz, Feng:2010gw}.

The total annihilation cross section, including all final states, is well defined, but the partial annihilation, scattering, and production cross sections with any specific standard model (SM) particles are model-dependent.  Measurement of any of these cross sections would dramatically constrain WIMP models and eliminate more exotic possibilities.  

There are limits from straightforward searches for astrophysical fluxes of annihilation products, direct nuclear scatterings in underground experiments, and collider events with missing energy.  Each of these searches has different underlying uncertainties and one technique may be much more sensitive than others if the WIMP framework is different than commonly supposed.  A convincing WIMP discovery will require observations by multiple experiments with different techniques that give a consistent picture. To gain detailed knowledge of WIMP properties and distributions, observations in multiple channels with comparable sensitivities are needed.

Searches for high-energy neutrinos from the Sun give strong limits on WIMP-nucleon scattering.  WIMPs passing through the Sun may rarely scatter with nuclei and become gravitationally bound.  Scattering can occur by coherent spin-independent~(SI) or valence spin-dependent~(SD) interactions.  Further scatterings thermalize WIMPs in the solar core, where they annihilate.  Only neutrinos can escape and potentially be detected.  When the capture rate, $\Gamma_{\rm C}$, and the annihilation rate, $\Gamma_{\rm A}$, are in equilibrium, as expected, an upper limit on the neutrino flux sets an upper limit on the WIMP-nucleon scattering cross section.

The most interesting limits using searches for high-energy neutrinos from the Sun are on SD WIMP-proton scattering, $\sigma_{\chi p}^{\rm SD}$.  Though the searches are based on the annihilation process, these limits are independent of $\langle \sigma_A v \rangle$, except for assumptions about the annihilation final states, which govern the detectability of the neutrinos.  High-energy neutrinos come from the few annihilation products that decay promptly, before losing energy in the solar medium, giving continuum spectra up to $E_\nu \sim m_\chi$ (direct annihilation to neutrino pairs, helicity-suppressed in many models, produces a line at $E_\nu = m_\chi$).  Strong limits on the SD WIMP-proton scattering cross section have been derived for large $m_\chi$ and certain final states, such as $W^+W^-, \tau^+\tau^-,$ and $b \bar{b}$.

Generalizing to other final states and a broader range of masses is challenging.  The number of high-energy neutrinos per annihilation is small ($N_\nu \sim 1$ for favorable final states) and their spectrum depends on the unknown final state.  Neutrino signal detection and atmospheric neutrino background rejection become easier for larger $m_\chi$, though neutrinos with $E_\nu \gtrsim 100$~GeV are significantly attenuated by interactions in the Sun.

What about the more common but seemingly less favorable final states?  As is well known, most final states ultimately produce pions and muons, which quickly lose energy and decay at rest, producing only MeV neutrinos, which have long been considered undetectable (e.g., see Ref.~\cite{Gaisser:1986ha, Ritz:1987mh} and many subsequent papers).

We propose a new probe of SD WIMP-proton scattering in standard solar WIMP capture scenarios.  First, we show that the pion yield from WIMP annihilation in the Sun -- produced directly in hadronic decays and further through inelastic interactions in the dense medium -- is large, relatively model-independent, and increases as $N_\pi \propto m_\chi$.   Second, we show that the subsequent low-energy neutrinos are much more detectable than previously thought, due to their high yield ($N_\nu \gg 1$) and known spectra, the low atmospheric neutrino backgrounds at low energy, and advances in detectors.  The advantages of this method are sensitivity to low WIMP masses and near-insensitivity to the choice of final state.  While the signal in high-energy neutrinos could vanish for unfavorable annihilation channels, for example if WIMPs annihilated only to light quarks, the low-energy signals would not. 

In the following, we review solar WIMP capture and annihilation, calculate the pion and neutrino yields per annihilation, estimate the signal and background rates in Super-Kamiokande (Super-K)~\cite{Fukuda:2002uc}, and derive new constraints on the SD WIMP-proton cross section.  We conclude by discussing likely improvements and the importance of complementary methods to allow for possible surprises in the astrophysics or physics of dark matter.  We give several examples of non-standard scenarios that highlight the importance of achieving this complementarity through a variety of experiments with comparable sensitivity.


\begin{figure}[t]
\includegraphics[width=3.25in]{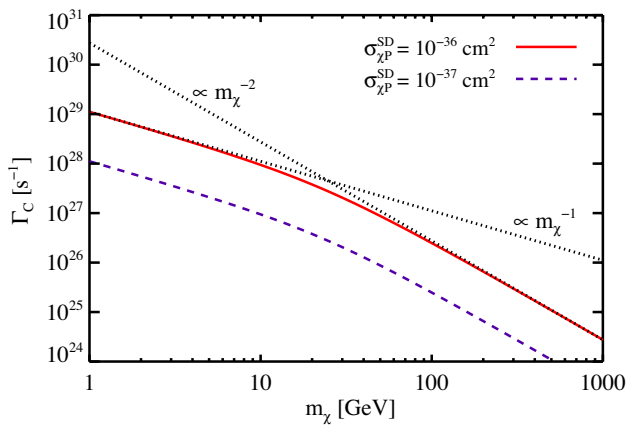}
\caption{Solar WIMP capture rate as a function of WIMP mass for two values of the SD WIMP-proton scattering cross section.  Evaporation (see text) is not included.
\label{fig:caprate}}
\vspace{-1.\baselineskip}
\end{figure}

\section{WIMP Capture and Annihilation}
When the probability of a WIMP scattering with a nucleus while passing through the Sun is small, the capture rate scales linearly with $\sigma_{\chi p}^{\rm SD}$ (we consider just the process of SD WIMP-proton scattering; if WIMPs also capture by SI scattering, that would only increase the neutrino detection rates calculated below ).  It also scales linearly with the number density, $\rho_{\chi}/m_\chi$, where $\rho_{\chi}$ is the local mass density of WIMPs.  For large masses, the capture rate falls more rapidly than $1/m_\chi$, due to kinematic suppression of the energy loss~\cite{Gould:1992, Peter:2009mk}.

Figure~\ref{fig:caprate} shows $\Gamma_{\rm C}$ computed with DarkSUSY~\citep{Gondolo:2004sc}.  We use the defaults of $\rho_{\chi}$ = 0.3~GeV/cm$^{3}$, a Maxwellian velocity distribution for WIMPs with a 3-D velocity dispersion of 270~km/s, and a circular velocity of the Sun of 220~km/s.  For the parameters considered here, the annihilation rate easily reaches equilibrium with the capture rate, so $\Gamma_{\rm A} = \Gamma_{\rm C}/2$.  We use DarkSUSY only to calculate the capture rate.  We treat the annihilation processes through a dedicated separate simulation, as described below.  Scattering interactions in the hot solar core can unbind WIMPs (``evaporation")~\cite{Gould:1987ju}, and this process is not included in DarkSUSY.  Detailed calculations show that it is important for low WIMP masses but negligible for larger masses.  To be conservative, we present our final results for only $m_\chi \gtrsim 4$~GeV, though less stringent limits could be calculated for lower masses. The effect of evaporation has recently been reinvestigated in the literature and we refer the reader to this discussion~\cite{Busoni:2013kaa}.


\begin{figure}[t]
\includegraphics[width=3.15in]{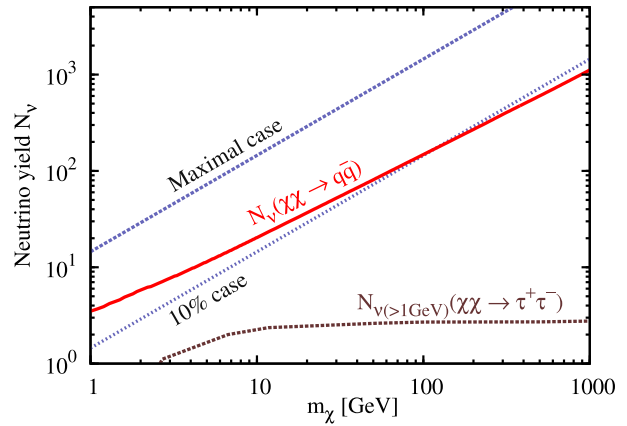}
\caption{Low-energy neutrino yield (summed over flavors) per WIMP annihilation ($\sqrt{s}=2 m_\chi$) in the solar core, obtained by simulating pion-induced hadronic showers in the solar medium.  Reference lines are shown for the cases where all or only 10\% of the annihilation energy goes into producing pions.  The high-energy neutrino yield for $\tau^{+}\tau^{-}$, relevant for current searches, is also shown.
\label{fig:pionyields}}
\vspace{-1.\baselineskip}
\end{figure}

\section{Pion Yield from Annihilation}
We calculate the number of pions produced per annihilation as a function of WIMP mass.  This depends on (i) the fraction of the annihilation energy in final states with eventual hadronic content, (ii) the hadronization of quarks and gluons, and (iii) all decay, interaction, and energy-loss processes in the solar medium.  The first is easily parameterized, the second is the same as in vacuum, but the third requires a detailed simulation.

For hadronic final states, we adopt results measured at ${\rm e}^{+}{\rm e}^{-}$ colliders~\cite{Nakamura:2010zzi}:
the initial pion multiplicity from hadronization is $N_\pi^{\rm initial} \simeq 3 + 4.5 \log_{10}^2(\sqrt{s}/{\rm GeV})$, pions dominate over other hadrons by an order of magnitude, and there are comparable populations of $\pi^{+}$, $\pi^{-}$, $\pi^{0}$.  WIMP self-annihilations and electron-positron collisions both have initial states with no quantum numbers or hadrons.  In vacuum, hadrons simply decay, producing a modest number of additional pions.  In matter, by contrast, the number of pions can be greatly amplified through hadronic interactions in the medium.

For high-energy charged pions, the hadronic interaction length in the Sun is short compared to the decay and continuous energy-loss lengths, so the number of pions increases in each generation of the hadronic shower until  loss processes dominate at low pion energies.  Neutral pions only decay, due to their short lifetime, diverting $1/3$ of the hadronic energy in each generation into an electromagnetic shower.  Charged pions are eventually brought to rest.  Negative pions are Coulomb-captured into pionic atoms with nuclei beyond hydrogen and then absorbed, and so do not produce neutrinos~\cite{Ponomarev:1973ya}.  Positive pions decay, producing 3 neutrinos with energies up to $E_\nu = 52.8$~MeV, through $\pi^+ \rightarrow \mu^+ \nu_\mu \rightarrow e^+ \nu_e \bar{\nu}_\mu \nu_{\mu}$.  

We used GEANT4~\cite{Agostinelli:2002hh} to model hadronic showers in the solar core, defined by a homogeneous volume with density 160~g/cm$^{3}$ and solar elemental abundances~\cite{Bahcall:1992hn}.  Each annihilation releases $\sqrt{s} = 2 m_\chi$ of energy, which we assume goes to a hadronic final state with the initial pion multiplicity $N_\pi^{\rm initial}$ and equal numbers of $\pi^+$, $\pi^-$, $\pi^0$.  We inject these into the solar core and calculate the resulting final pion yield $N_\pi$.  We find that our results are not sensitive to reasonable variations in the density, composition, or ionization state.  Variations in the density by an order of magnitude did not change the neutrino yields in our simulation, nor did large increases in the abundance of heavier elements; both variations are much larger than the uncertainties in the standard solar model.  At these energies, pion energy losses on photon targets are negligible~\cite{Dermer:2009zz}.

Figure~\ref{fig:pionyields} shows our results for the low-energy neutrino yield, $N_\nu = 3 N_{\pi^{+}}$, per WIMP annihilation.  The near-linear scaling of $N_{\nu}$ with annihilation energy indicates the primary importance of the total hadronic energy, with the mild deviations reflecting losses due to $\pi^{0}$ production.  Comparable results are indeed obtained when the initial number of pions is varied at fixed injection energy.  Because of pion dominance in hadronic showers, similar results are expected for arbitrary hadronic annihilation products, and this was confirmed in some representative cases.  Our calculated results are comparable to simply assuming that $\sim 10\%$ of the annihilation energy goes into producing pions, i.e., $N_\pi \sim 0.1\, (2 m_\chi/m_\pi)$.  This is similar to the fraction of energy in hadronic showers in ultra-high-energy cosmic ray interactions in Earth's atmosphere~\cite{Beatty:2009zz}; the details differ, but the common element is that the dominant charged-pion loss process is hadronic scattering.

For low-energy neutrinos, the rate and spectrum are very similar for nearly all final states, unlike the case for high-energy neutrinos.  Quarks, gluons, weak bosons, and tau leptons all decay dominantly into hadrons, which produce pions.  Muons also produce low-energy neutrinos.  For both pions and muons, the decays happen after the particles have been brought to rest, and the shapes of their decay spectra are well known.  High-energy neutrinos are covered in current searches.  For gamma rays and electrons, there are very few neutrinos produced (however, those final states are well probed by searches for annihilation fluxes from Galactic and extragalactic dark matter structures).


\section{Neutrino Signals and Backgrounds}
The neutrino flux is $\phi_\nu = N_\nu \Gamma_A / (4\pi D^2)$, where $D$ is the Earth-Sun distance.  The neutrino spectra are well known: $\pi^+ \rightarrow \mu^+ \nu_\mu$ gives a line spectrum for $\nu_\mu$, and $\mu^+ \rightarrow e^+ \nu_e \bar{\nu}_\mu$ gives a softer continuum spectrum for $\nu_e$ and a harder one for $\bar{\nu}_\mu$~\cite{Nakamura:2010zzi}.  The Sun and Earth are transparent to these neutrinos.  We consider detection in Super-K, using the interaction $\bar{\nu}_e p \rightarrow e^+ n$.  Matter-enhanced mixing in the Sun gives a flavor-change probability of $P(\bar{\nu}_\mu \rightarrow \bar{\nu}_e) \simeq 1/6$ for either neutrino mass hierarchy~\cite{Lehnert:2007fv}.

The detectability considerations are similar to those for the diffuse supernova neutrino background, though that peaks at lower energies~\cite{Beacom:2010kk, Bays:2011si}, and for monopole-catalyzed proton decays in the Sun, though that has lower pion injection energies~\cite{Ueno:2012md}.  For this neutrino interaction~\cite{Vogel:1999zy, Strumia:2003zx}, the cross section rises as $\sigma_\nu \sim 10^{-43} \, (E_\nu / {\rm\ MeV})^2 {\rm\ cm}^2$ and the positron is detectable; it has nearly the full neutrino energy, but only a weak forward anisotropy in this energy range.  Super-K has $1.5 \times 10^{33}$ free (hydrogen) protons in the 22.5~kton of water in the fiducial volume.

\begin{figure}[t]
\includegraphics[width=3.25in]{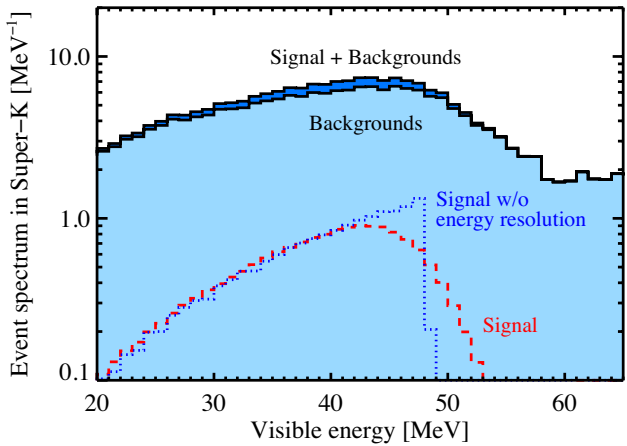}
\caption{Detectable signal in Super-K induced by low-energy $\bar{\nu}_e$ from solar WIMP annihilation, along with the measured background (both 4.1 livetime years).  The signal shape is independent of WIMP properties, and its normalization scales with $\sigma_{\chi p}^{\rm SD}$ (here chosen to be at the edge of exclusion). For a WIMP mass of 10~GeV, the signal shown corresponds to a cross section of $4.5\times 10^{-37} {\rm cm}^2$. 
\label{fig:sigback}}
\vspace{-1.\baselineskip}
\end{figure}

Figure~\ref{fig:sigback} shows the observable $\bar{\nu}_e$ signal spectrum, accounting for neutrino mixing, the full neutrino cross section and kinematics, and detector energy resolution~\cite{Fukuda:2002uc}; the detection efficiency is near unity~\cite{Bays:2011si}.  For simplicity, we consider just the 4.1 livetime years of the Super-K-I phase (1996--2001).  The measured background spectrum, which has a small absolute rate, is also shown; its components are displayed in Ref.~\cite{Bays:2011si}.  The largest component is electrons and positrons from the at-rest decays of sub-\v{C}erenkov (``invisible") muons produced by atmospheric neutrinos.  The spectrum of this background is the same as the signal $\bar{\nu}_\mu$ (which mixes to $\bar{\nu}_e$) flux spectrum.  However, the detection spectrum of the signal is weighted with the energy-dependent neutrino interaction cross section,  changing its shape.

\section{Constraints on WIMP Properties}
Figure~\ref{fig:sens} shows the Super-K sensitivity to $\sigma_{\chi p}^{\rm SD}$ with present data.  We estimate this by first calculating the expected numbers of signal and background events, summed over the energy range 25--50~MeV.  These numbers are 15 and 135 for the case shown in Fig.~\ref{fig:sigback}.  For each WIMP mass, we estimate the 90\% C.L.~upper limit on $\sigma_{\chi p}^{\rm SD}$ that would be set if no signal were seen and statistical fluctuations in the background were taken into account. An experimental search will require a more sophisticated treatment. However, the search is very feasible and similar Super-K searches in the same energy region have demonstrated this~\cite{Bays:2011si,Ueno:2012md}. Individual background component spectra can be understood and normalized by analyzing the data over a larger energy range than we use. Our approach is conservative, and improvements are discussed below.

Our estimate is compared with published upper limits from direct nuclear scattering experiments (see also new results in Ref.~\cite{Behnke:2012ys}) and indirect searches for high-energy neutrinos from the Sun.  The sensitivities of both are expected to improve greatly with better detectors and analyses.  However, it is very difficult for each to extend to lower $m_\chi$, as the signals decrease while the backgrounds increase, and there are kinematic thresholds.  In contrast, the sensitivity of our proposed low-energy neutrino signal improves with decreasing $m_\chi$.  There are also strong limits on the SD WIMP-nucleon scattering cross section that can be deduced from limits on mono-jet and mono-photon signals at hadron colliders; these require that the masses of mediators coupling WIMPs to the SM be large~\cite{Goodman:2010ku, Bai:2010hh}.


\section{Conclusions}
We propose a new probe of the SD WIMP-nucleon scattering cross section, using low-energy neutrinos produced through pion multiplication and decay following WIMP annihilations in the Sun.  We estimate the prospects for Super-K, finding sensitivity to $\sigma_{\chi P}^{\rm SD}$ in a range competitive to that of other, rather different, experiments.  Importantly, our results are nearly insensitive to the annihilation final states and the details of the astrophysical inputs (in standard scenarios~\cite{Rott:2011fh}); non-standard scenarios can be different and are mentioned below.  In addition, the sensitivity easily extends to the region of low masses, which is currently of great interest and is challenging to probe with other methods.  A dedicated study by the Super-K collaboration, using more data, neutrino-electron scattering signals for all flavors, and full energy spectra and angular distributions, should give immediate improvements over our estimates.

\begin{figure}[t]
\includegraphics[width=3.25in]{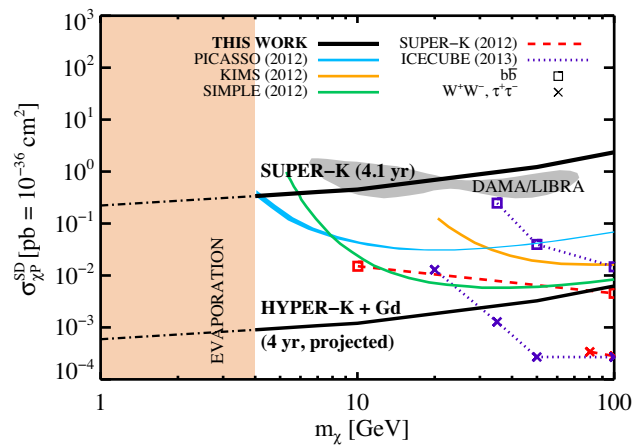}
\caption{New sensitivity to $\sigma_{\chi p}^{\rm SD}$ using low-energy $\bar{\nu}_e$ in Super-K.  Possible improvements using neutron tagging with Gadolinium (details see text) are included in the projected sensitivity for 4~years of Hyper-Kamiokande~\cite{Abe:2011ts} data.  For $m_\chi \lesssim 4$~GeV, evaporation of solar WIMPs will degrade the signal.  Published upper limits (90\% C.L.) from direct scattering~\cite{Archambault:2012pm, Kim:2012rz, Felizardo:2010mi} and high-energy neutrino searches are shown~\cite{Tanaka:2011uf, Aartsen:2012kia}, along with the possible DAMA/LIBRA signal region~\cite{Bernabei:2010mq, Savage:2008er}. 
\label{fig:sens}}
\vspace{-1.\baselineskip}
\end{figure}

In the future, Super-K may be enhanced with dissolved gadolinium to allow neutron detection and thereby better separation of signals and backgrounds~\cite{Beacom:2003nk}.  If a combination of techniques removed the backgrounds, then the sensitivity to the neutrino signal and hence also $\sigma_{\chi P}^{\rm SD}$ could improve by $\sim 15$, beyond which even Super-K is too small to expect any signal events.  The quoted improvement factor is obtained with inverse beta decay alone assuming a signal event detection in a zero background environment. Hyper-Kamiokande~\cite{Abe:2011ts}, which is intended to be about 25 times larger than Super-K and which also may have gadolinium, could potentially improve on present estimates by up to $\sim 15 \times 25 \sim 375$. Figure~\ref{fig:sens} shows our rough Hyper-Kamiokande sensitivity estimate under the assumption that backgrounds can be reduced and that only the inverse beta decay detection channel is utilized. Further studies on how to reduce backgrounds as well as contributions from other detection channels are needed but are beyond the scope of this paper. Proposed large liquid scintillator~\cite{Wurm:2011zn} or liquid argon~\cite{Rubbia:2009md} detectors would also have interesting sensitivity.  

Direct, indirect, and collider probes of the SD WIMP-proton scattering cross section rely on different assumptions and hence are complementary.  Multiple methods with comparable sensitivity are needed to test results from one search against the others.  These tests could provide deep insights into the astrophysical distributions and particle properties of WIMPs.

If common assumptions about dark matter are incorrect, then the relative power of different methods could change dramatically.  A dark matter disk with low-velocity WIMPs could enhance both types of neutrino signals~\cite{Bruch:2009rp}.  A time-varying WIMP flux from dark matter substructures could alter the relationship between neutrino and direct nuclear scattering signals~\cite{Koushiappas:2009ee}.  WIMP annihilations through a new low-mass force carrier that decays only into low-mass SM particles~\cite{ArkaniHamed:2008qn} could have vanishing high-energy but strong low-energy neutrino signals.  These are just a few examples.  Most generally, a combination of experiments will be required to reconstruct or constrain the complete couplings of the SM to dark matter, whether WIMPs or something more exotic.

\medskip
{\bf Note added:}
As this work was being completed, we learned of an independent study with similar results by Bernal, Martin-Albo, and Palomares-Ruiz~\cite{Bernal:2012qh}. Both works were submitted simultaneously to arXiv.


\begin{acknowledgments}
%
We are grateful to Basu Dasgupta, Louie Strigari, and Petr Vogel for helpful discussions.  J.S.-G. was supported by NASA through Einstein Postdoctoral Fellowship grant PF1-120089 awarded by the Chandra X-ray Center, which is operated by the Smithsonian Astrophysical Observatory for NASA under contract NAS8-03060.  J.F.B. was supported by NSF Grant PHY-1101216.
\end{acknowledgments}


\end{document}